# Design, fabrication, and spectral characterization of TM-polarized metamaterials-based narrowband infrared filter

*Golsa Mirbagheri[1], David T. Crouse[2]*

*Duke University Pratt School of Engineering[1]*
*Computer and Electrical Department of Clarkson University[2]*

**ABSTRACT**

Hyperbolic Metamaterials, as a non-magnetic anisotropic artificial structure, show metal properties in one direction and dielectric behavior in orthogonal directions. The proposed hyperbolic metamaterial filter in this project is designed with the metal wire mesh perpendicular to the alternative layers of dielectric materials, keeps TM center wavelength unchanged for the different angle of incident light in MDIR regime. The geometric size of this nanostructure is smaller than the working wavelength and supports big wavevectors due to hyperbolic dispersion. In contrast with conventional Bragg stack, the copper fakir bed makes the transmission properties of the filter the same. For this purpose, the state-of-the-art fabrication methods are required to make such small dimensions in alternative layers of amorphous silicon and silicon dioxide. In this work, first we demonstrate the simulation of Bragg stack with RCWA and finite element methods. Then we focus on our first-time multistep lithography method used to fabricate the filter at Cornell University's Nanoscale Science and Technology Center. Finally, we experimentally verify the optical characteristic of the fabricated filter using Fourier-transform infrared spectroscopy. The experimental and spectrometry data shows that transmission properties of the hyperbolic metamaterial filter remain the same for oblique TM polarized incident light.

**Keywords:** nanofabrication, hyperspectral metamaterial, lithography, angle independency, negative photoresist

## 1. INTRODUCTION

Advance in nanofabrication techniques facilitated building the different types of the nanostructures in the past decades. Among various types of the metamaterials, hyperbolic metamaterials (HMM) have gained more attentions in sensing and imaging applications [1, 2]. The dielectric tensor of this media has positive and negative values, makes the non-magnetic HMM anisotropic. In our work, the proposed HMM Brag stack is composed of metal wire mesh (as hyperbolic metamaterial) and conventional Bragg stack, makes the angle shift negligible and therefore, remains the TM center wavelength the same. This is in contrast with angle-dependent conventional notch filter that center wavelength of light changes for different angles of incident light. This type of filter is applied in many hyperspectral and imaging systems that are sensitive to the shift of the wavelength of light. The unit cell of the homogenous HMM is much smaller than the wavelength of light. The shrinkage of feature size of these structures leads to investigate on new techniques of lithography, as the old fashion fabrication methods have the resolution limits [3]. Therefore, novel fabrication methods have been introduced to guarantee the high resolution and mass production of nanostructures. The main goal of this article is to introduce the new fabrication methods to build the proposed metal-dielectric HMMS and then measuring the optical properties of the filter, in addition to show the simulation results with RCWA and finite element methods. In first section, the background and related work to HMM is explained. Section 2 investigates on the fabrication technology used to construct the proposed filter. Next section discusses about the optical result of the fabricated HMM and the last section, concludes the project.

---

[1] Send correspondence to Golsa Mirbagheri
E-mail: gm226@duke.edu

## 2. BACKGROUND

**Related Works**

Hyperbolic metamaterials were originally manufactured for the purpose of diffraction limit in imaging applications. However, recently they have been investigated intensively in opto-electronics area for their extraordinary phenomena including the high-k modes support, phase transition and density of photonic states [4, 5, 6]. In this part, some recent works relative to hyperbolic metamaterials and surface plasmon are revied.  In [7] , the surface plasmon dispersion is adjusted by tuning the permittivity and electric filed of the metal-like multilayers structures, this is applicable by adding several dielectric layers or doped semiconductor (conductive) layers. The bull's eye subwavelength apertures are explained in [8]. These SP structures are similar to antenna, in such a way that their concentric periodic circles couple the light cone with SP modes and make a rise to the field around the aperture, which leads to extraordinary transmission. In [9] showed that the engineered structure with dimensions smaller than working wavelength can have the negative refraction, like negative index metamaterial (NIM). The dispersion relation of the proposed indefinite hyperbolic metamaterial media is composed of perpendicular negative and parallel positive permittivities. The pointing vector in this structure is normal to the anisotropic surface, but in different direction in isotropic media, which is in contrast with NIM with anti-parallel directions. High index prism or periodic structure were used to excite surface plasmon modes with light cone, showed in [10]. In [11], the pixelated guided-mode resonant filter (GMRF) is discussed to calculate the $CO_2$ and $H_2O$ absorption bands in MWIR. The narrowband hyperspectral GMRF acts as a Fabry-Perot microcavity, mirrors the pixels of the filter to the sensor pixels. In [12], the angle-independent filter composed of high refractive index material is described. The Fabry-Perot cavity-based filter has the a-Si dielectric layer sandwiched between the silver layer and chrome layer, makes the considerable reflection phase shift that reduce the angle sensitivity [13, 14]. In [15], the bandpass Fabry-Perot resonator filter composed of two high index Bragg reflectors is separated by a low loss dielectric subwavelength layer. By changing the geometry properties of the middle meta-surface layer, the center wavelength is controlled as phasing shift function. The angle independent etalon resonator in [16], composed of a high index $TiO_2$ cavity layer sandwiched between two Ag/Ge mirrors, shows more polarization and angle independency in comparison with the filter with low index $SiO_2$ sandwiched between metal reflectors.

**Conventional Bragg Stack Filters**

The conventional Bragg stack consists of alternative layers of materials with high and low dielectric constant.  To have the constructive interference in the stopband Bragg stack filter, the thickness of a layer (usually middle layer) should be twice as the other layers, namely $t_i = \lambda_c/2n_i$, where $n_i$ is the index of refraction of layer i and $\lambda_c$ is the narrow passband center wavelength of incident light. The resonant condition, which happens in this middle layer, is written as Eq 1 [17, 18, 19].

$$2k_{j,z}t_j = 2n_jk\cos(\theta_j)t_j = 2\pi m \qquad (1)\ [18]$$

In Eq 1, the phase of narrow passband center-wavelength light is an integer of $2\pi$, where $k_{j,z}$ is the wave vector, k is $2\pi/\lambda_o$ and $n_j = \sqrt{\varepsilon_j}$. However, the notch filter performance restricts to the angle of incidence light, such that the center wavelength of light changes for off-normal incidence light [20, 21, 22, 23]. The permittivity tensor of anisotropic uniaxial non-magnetic is composed of elements $\varepsilon_{zz}$ and $\varepsilon_{xy}$ derived from Maxwell-Garnett theory, where $N$ is the filling ratio, $\varepsilon_w$, is the wire permittivity and $\varepsilon_d$ is the dielectric permittivity in Eq 2 [18].

$$\varepsilon_{xy} = \frac{\varepsilon_d^2(1-N)+ \varepsilon_d\varepsilon_w(1+N)}{\varepsilon_d(1+N)+ \varepsilon_w(1-N)} \qquad \varepsilon_{zz} = \varepsilon_d\ (1-N) + \varepsilon_w\ N \qquad (2)\ [18]$$

In Eq 3 the dispersion relation for TM incident light is given [24, 17, 25, 26]. If condition $|\varepsilon_{xy}|<<|\varepsilon_{zz}|$ holds, $k_x^2(\theta)$ would be negligible and this makes $k_z$ insensitive to the angle of incident light, shown in Eq 4 [27, 28, 29, 30]. The condition $|\varepsilon_{xy}|<<|\varepsilon_{zz}|$ enables the dispersion-less filter to operate in the shortwave and midwave infrared regions.

$$k_0^2 = \frac{k_z^2}{\varepsilon_{xy}} + \frac{k_x^2}{\varepsilon_{zz}} \qquad (3) [18]$$

$$k_z(\theta) = \sqrt{\varepsilon_{xy} k_0^2 - \frac{\varepsilon_{xy}}{\varepsilon_{zz}} k_x^2(\theta)} \approx \sqrt{\varepsilon_{xy}} k_0 \approx k_z(\theta = 0) \qquad (4) [18]$$

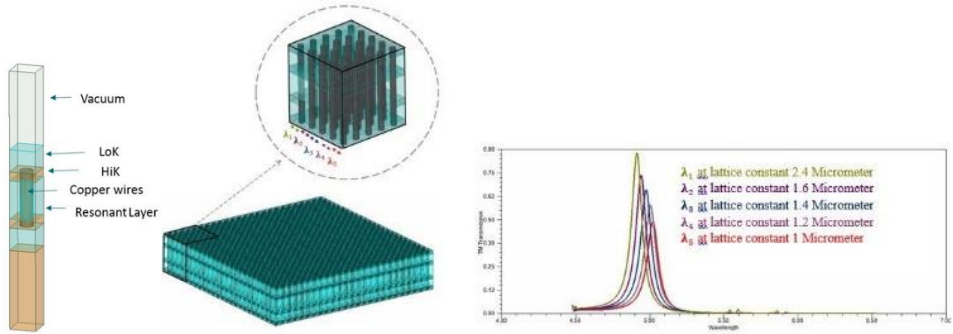

Figure 1. In Left, the simulated HMBS with unit cell 1µm in HFSS. In middle, each pixel of the HMBS filter with different geometry, transmits different wavelength of light. In right, center wavelength changes by modifying the unit cell geometry. [18]

Figure 1 (left) shows the simulated proposed filter with finite element method, where diameter of wire is 0.5 and the lattice constant is *1µm*. The HMBS is composed of alternative dielectric layers of silicon oxide and amorphous silicon, with subwavelength-sized copper wires etched in the three middle layers of the stack. The thickness of SiO2 and a-Si dielectric layers are 880*nm* and 330*nm* sequentially, while the middle resonant layer thickness is 1.5*µm*. Due to fabrication issues, it was not possible to etch the whole 7 layers down; also, new methods were devised to electroplate the copper wires in the middle layers. The HMBS can transmit the different wavelength of light by adjusting the size of each unit cell, as depicted in Figure 1 (middle). The transmitted light then is broken apart and detected by relative pixel of detector array in spectrometer. The resonant condition in the proposed HMBS happens when filling ratio (N) changed by modifying unit cell size [31, 32, 33, 34]. This leads to change in $\frac{\varepsilon_{xy}}{\varepsilon_{zz}}$ in Eq 4 and consequently, center wavelength changes based on adjusting $k_z$ in Eq 3 [35, 36, 37, 38]. In hyperspectral imaging applications, the notch filters need collimated light to keep the dispersion properties unchanged. This adds extra optic equipment, cost and wight to the imaging systems. In this work, the subwavelength-sized wire mesh, perpendicular to the layers of HMBS, removes the dependency of angle of light and therefore, the filter is applicable for both focused and collimated incident MIDIR light [39, 40, 41, 42].

### 3. Full-Wave Finite Element Modeling Simulation

The Maxwell-Garnett theory applies effective media approximation approach to calculate the permittivity, as discussed before [43, 44, 45]. However, in finite element method, the frequency-dependent epsilon of the materials is computed which resulted in more reliable result. Therefore, the optical results of these two methods are different. Figure 2 shows different results between Maxwell-Garnett and full-wave electromagnetic modeling. Furthermore, the electric filed is close to the metal wires in Maxwell-Garnett theory, while uniform filed is distributed around copper wire in FEM modeling [19].

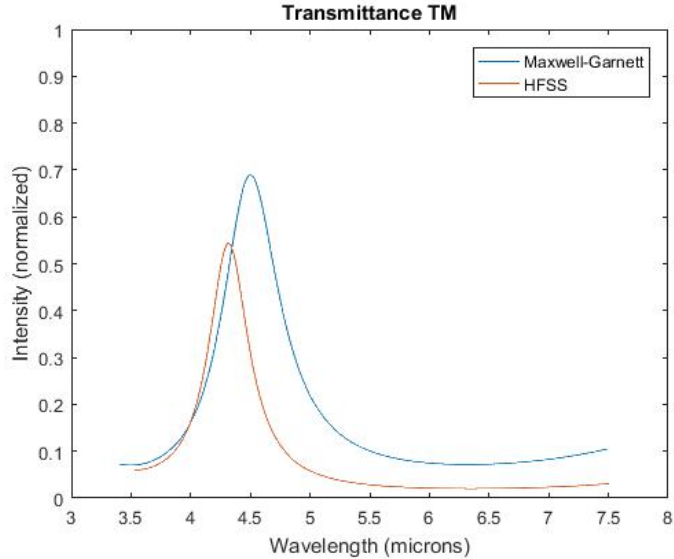

Figure 2. Comparison of the optical properties of HMBS simulated by Maxwell-Garnett theory and HFSS [19]

## 4. FABRICATION

The Bragg stack composed of alternative layers of a-Si as high dielectric and SiO2 as low dielectric including the copper cylinders was fabricated by CMOS fabrication methods at Cornell University's Nanoscale Science and Technology Center (CNF). In fabrication process, there were lithography challenges to etch through the whole layers of the filter, in addition to the electroplating the copper in the etched holes. In this section, first all the steps to build up the chip are outlined, then we probe the novel techniques to overcome the fabrication issues.

### Deposit alternative layers

The first step of building the stack was depositing the dielectric layers of a-Si and SiO2 on 4inch silicon wafer. Depending on the fabrication plan, the whole layers could be deposited in one step or two with plasma-enhanced chemical vapor deposition (PECVD) method. Due to aspect ratio, etching the whole 3 middle layers in one step was not possible, so first two layers of a-Si and SiO2 were deposited on top the first SiO2 layer to be etched before depositing the fourth layer. After lithography process, the two layers etched, and the remaining photoresist removed. However, depositing the second a-Si on top of the two middle layers filled the holes, as shown in Figure 3. The $SiO_2$ was deposited with rate of 20sccm $SiH_4$ and 2500sccm $N_2O$ at 140w AC power and 18mTorr pressure in 300°C. The a-Si was deposited with rate of 25sccm $SiH_4$ and 475sccm Ar at 140w AC power and 3mTorr pressure in 200°C. The $SiO_2$ layers thickness is 880nm which is double in resonant layer thickness, while the a-Si layers have the thickness of 330nm deposited on 4inch 525µm double-side polished silicon wafer. Different scenarios were devised to etch down the three dielectric layers which are discusses in next section [19]. The next steps were using hard mask and backside alignment for the purpose of multiple steps lithography.

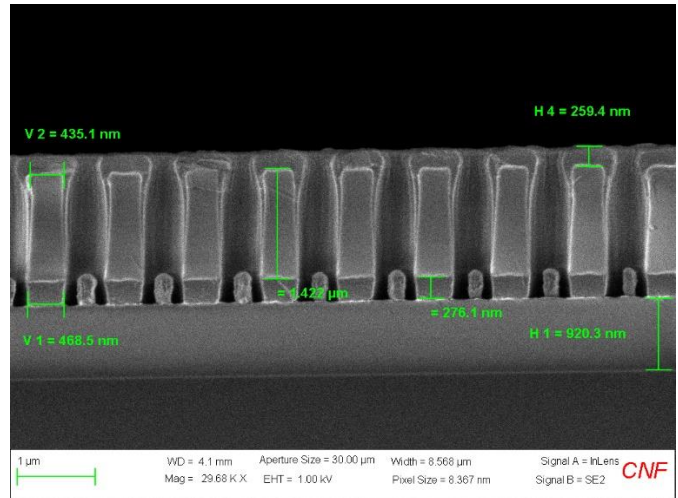

Figure 3. The top a-Si deposited on the etched two middle layers, left some a-Si in the holes.

### Photolithography

Nanolithography technology brought significant advance to fabrication of nanometer-scale structures [46, 47]. We started with positive photoresist which is spin-coated and then exposed on the wafer with ultraviolet light through a dark-field mask. The exposed photoresist resolved and disappeared by the solution in developing process, left the designed pattern on the wafer, as depicted in Figure 4 (left). This is different for negative photoresist, where unexposed area is resolved with the solution. Figure 4 (right) shows the lithography pattern on the wafer with negative resist and bright field mask. In this work both positive and negative photoresist used to etch down the middle layers; however, the desired etching achieved using the negative photoresist and bright field mask.

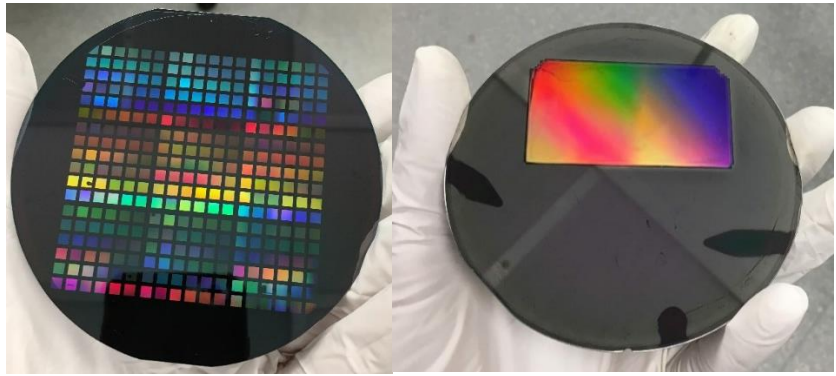

Figure 4. Lithography Matrix of different energy and dose with positive resist and dark field mask (left) negative resist and bright field mask (right) [19]

### Etching

Etching the several dielectric layers of Bragg stack was such a challenging work in the fabrication process. For this issue, different methods were applied including thicker photoresists, hard mask or separate lithography multi-steps. The reactive ion etching (RIE) method was used to etch the a-Si and $SiO_2$ layers of the filter. In this process, 18sccm $CH_2F_2$ RIE etching gas beside 72sccm Helium gas carved the $SiO_2$ layer down at pressure 4mTorr and temperature 10C, while for etching the a-Si layers 20sccm HBr gas at 15mTorr pressure and temperate16c was applied. The first challenge to etch the middle $SiO_2$ was interaction of $CH_2F_2$ gas with the a-Si layer on top of the resonance layer, which creates polymer in the holes and consumes the 600nm photoresist faster. Figure 5 shows the etching process stopped in the middle of resonant layer. This problem made us to use the thicker photoresist (1400nm), however because of aspect ratio thicker photoresist did not work as well.

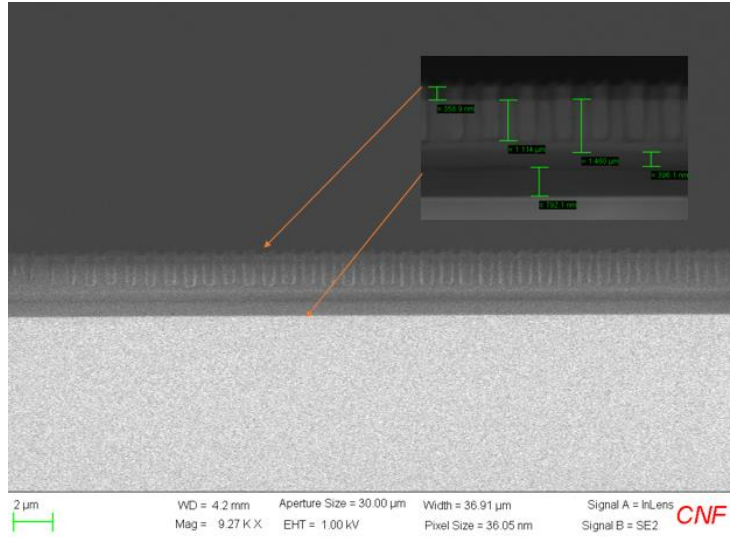

Figure 5. No photoresist left after etching the top a-Si and middle $SiO_2$ layers [19]

The other alternative way to etch the layers down was implementing the hard mask. The hard mask can be chrome or aluminum oxide, slowing running out of photoresist to etch the dielectric layers. First, Cr hard mask etched by 27sccm $Cl_2$, 1sccm $O_2$ and 2sccm Ar with RIE/ICP 5W/800W. The 110nm Cr hard mask on the top a-Si layer interacted with the HBr gas used for etching the a-Si layers, made polymers in the holes. Furthermore, the $Al_2O_3$ mask etched so fast in contact with $Ch_2F_2$ gas to etch $SiO_2$ layer. One potential method, not used in this project, is suggested to use both $Al_2O_3$ and Cr hard mask, such that Cr sandwiched between two $Al_2O_3$ hard mask layers to avoid interaction between etching gasses and dielectric materials. Other experiments performed to etch the middle resonant layer, including using $CHF_3O_2$ which created less polymer in contraction with $SiO_2$. However, the latest experiments with $CHF_3O_2$ gas made the side walls wider, as shown in Figure 6 [18].

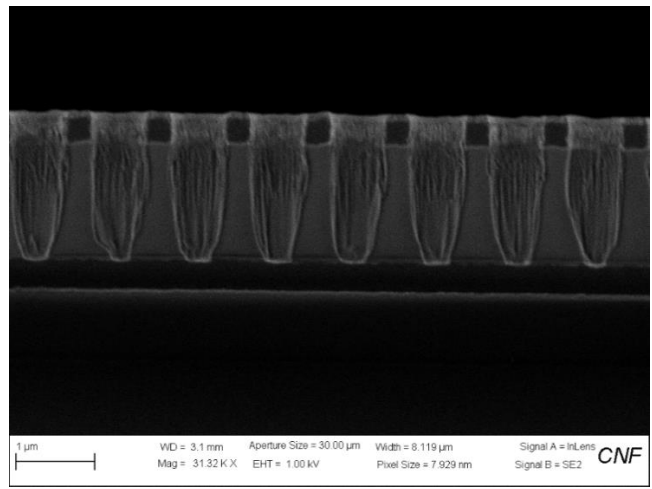

Figure 6. SEM view of middle $SiO_2$ layer etched by $CHF_3O_2$ gas

The latest trick to successfully etch down the middle dielectric layers was based on backside alignment in separate multi-steps lithography. At first, 4 marks was designed on back (or front) of the wafers during the lithography process, and then they were etched about few nm. These marks were used as alignment marks, such that in each lithography step the pattern is printed in the exact same place that was printed before by ASML lithography program. Therefore, the etching process is continued from where the features etched and resist-ashed last time. The misalignment in our work in less than 1nm, as shown in Figure 7.

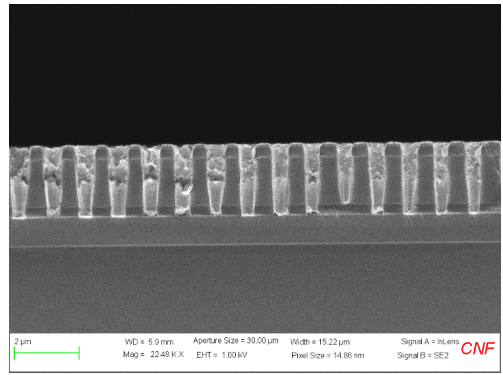

Figure 7. The three middle layers were etched, then electroplated with Cu and polished [19]

In the multi-step photolithography process, after second or third steps, the etched holes filled with so much positive photoresist. The stepper tool has the limit to focus into the depth of the etched holes and, high dose is needed for such a thick resist, which leads to increase the size of the features. For this problem, we used negative photoresist with bright-filed mask to keep the feature size the same [48]. The developer solution removes the unexposed negative photoresist on the wafer and therefore, the resist inside of the holes came out. This is in contrast with positive photoresist which exposed resist area are removed in developing step [49]. Figure 8 shows that by using the bright filed mask with negative resist, the same photoresist resist area on the wafer is resolvable when using positive resist and dark field mask [50].

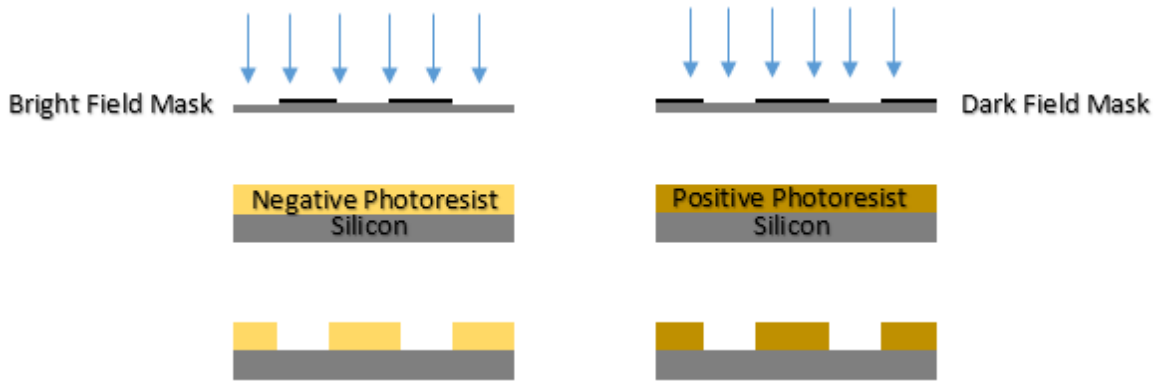

Figure 8: Bright filed mask with negative resist (left). Dark filed mask and positive resist (right)

**Electroplating**

In the next step, before starting the copper electroplating process, 5nm Ti and 80nm Pt sputtered on the side walls of the etched features. The conductive Pt conformally sputtered on the insulator side walls of the dielectric layers, as an adhesion layer, made copper ions stick to the edges of the holes. During the electroplating process, the copper sulfate solution splits into positive and negative copper ions, the positive sulfate copper ions in solution flew to the cathode electron and made the thin layer of copper sits on the filter [18, 51, 52, 53]. In this project, the electroplating of the filter was such a challenging work, so at first, we tried to test with the 2 layers filter including the SiO$_2$ layer deposited on wafer. The electroplating time was adjusted based on the current and concentration of the solution. The forward current set up for 120mA for 200ms (60ms on/20ms off), whereas reverse current was 350mA for 180ms (2ms on/178ms off), the whole process took 4 hours for copper ions conformally fill the whole features. Figure 9 shows the holes in 7-layers BS filled with copper completely, then the last layer SiO$_2$ deposited on the polished filter. The slurry used to polish the Cu was combination of M8540-P6 with hydrogen peroxide (14.8 to 1 part by volume). The abrasive particles in the slurry polished the copper particles up to the top a-Si layer, following the metal electrodeposition step. The last SiO$_2$ layer deposited on the polished wafer, as depicted in Figure 9. This worked was published in the NanoMeter newsletter of the Cornell Nano Scale Facility [54].

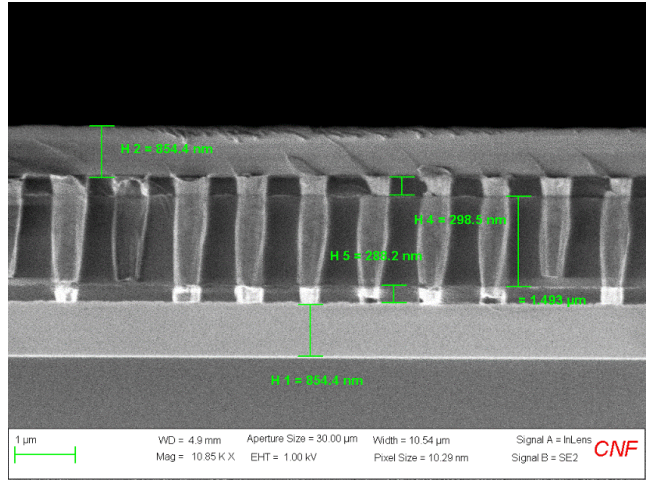

Figure 9. The SEM view of final fabricated HMBS at CNF [19]

## 5. EXPERIMENTAL TEST

In this project, the Fourier-Transform Infrared Spectroscopy (FTIR) technique was used to measure the infrared spectrum of the fabricated filter [55, 56]. The iS50 FTIR setup included the KBr beam splitter and nitrogen cooled DTGS detector with spot size of 100 × 100 μm$^2$. In order to collect spectral data for different angles of light, the IR light was collimated by different lenses and beam expander [18, 57]. Figure 10 shows a schematic of optical setting of light collimation at Clarkson University's lab. To reduce the noise and extra peaks of detected spectra, the entire optical setup was purged. For this purpose, the nitrogen gas tuned to the pressure of 20 psig and flew in the sealed box to eliminate the effect of the water vapor or carbon monoxide of FTIR chamber compartment and the sealed box.

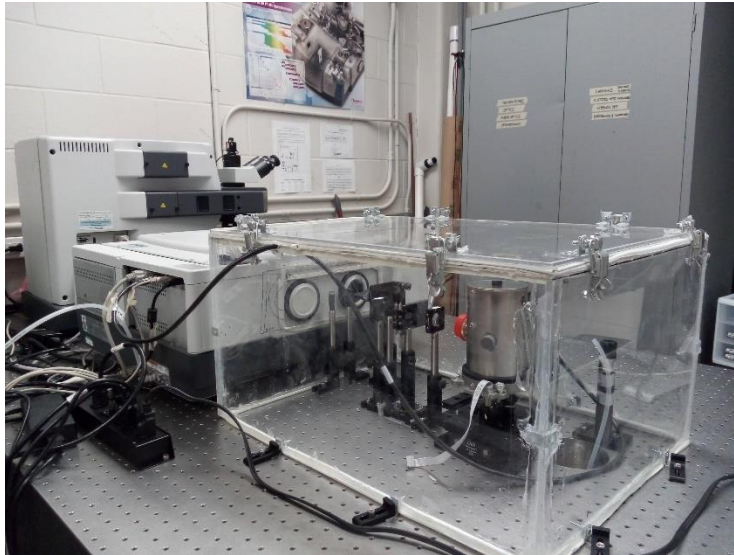

Figure 10. collimation setup and purging the system at Clarkson University lab [19]

Figure 11 shows the result of FTIR measured spectral data of fabricated filter. The MCT detector captured the TM transmission of the HMBS, show the 33% narrowband peak at the wavelength of 4.33μm. The result showed that for different angle of light, started from 0 to 25 degree by step 5, the center wavelength changed for few nanometers, or on the other hand, remained unchanged. The maximum band peak could have been narrower by increasing the number of the fabricated filter layers [18].

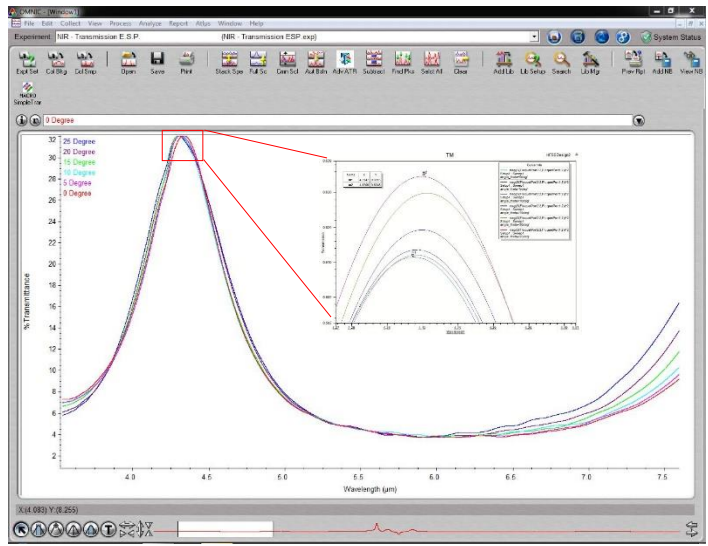

Figure 11. Locked center-wavelength of HMBS for angle sweep 0-25 degree measured in FTIR

The result of finite element method, Maxwell-Garnett theory and FTIR spectroscopy are compared in Figure 12. In Finite element method, frequency dependent permittivities were implemented to calculate the spectral result, more specific result captured which is different than Maxwell-Garnett theory where permittivities calculated based on $\varepsilon_{zz}$ and $\varepsilon_{xy}$. However, the lossy copper we electroplated to fabricate the proposed filter, made the transmission lower than the other two methods. It should be mentioned that the last step of CMP was done with hand, and therefore without performing CMP brushing, few picometer of Cu left on the surface of the filter that led to reduce the transmission. However, based on the result shown in Figure 11, the center wavelength of filter shows angle-independency for TM light, which indicated that the final main of this project is achieved.

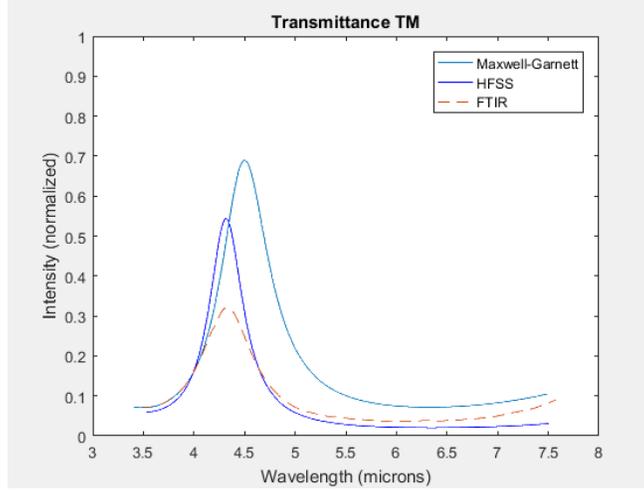

Figure 12. Transmission of Maxwell-Garnet theory, finite element method and FTIR implemented in MATLAB.

## 6. CONCLUSION

In this work, we combined the traditional notch filter with hyperbolic metamaterial to reduce the dependency of the narrowband TM transmission peak to normal and oblique TM incident light. The simulation and analysis of the TM polarized angle-independent HMBS were reviewed with both Maxwell-Garnett theory and finite element method in MWIR regime. By fabricating the Cu nanocylinders perpendicular to the dielectric layers of $SiO_2$ and a-Si of the filter, the center wavelength of the HMBS showed negligible shift to shorter wavelength for different angles of incidence light. One of the novel fabrication methods were used to build the filter was the multistep separate lithography process, made it possible to etch down the middle dielectric layers. The HMBS filter can be applicable for different applications of SWIR, MWIR, LWIR by adjusting the dimension size and materials of the filter.

## 7. ACKNOWLEDGEMENT

This work was performed in part at the Cornell NanoScale Facility, a member of the National Nanotechnology Coordinated Infrastructure (NNCI), which is supported by the National Science Foundation (Grant NNCI-2025233). We would like to thank the NSF Industry/University Cooperative Research Center for Metamaterials to support this project.